\documentclass[aps,pra,nofootinbib,preprint,amsmath,amssymb,floatfix]{revtex4}
\usepackage{mathtools}
\usepackage{color}
\usepackage{graphicx}
\usepackage{comment}

\begin{document}

\newcommand{\tc}{\textcolor}
\newcommand{\g}{blue}
\newcommand{\ve}{\varepsilon}
\DeclarePairedDelimiter\abs{\lvert}{\rvert}
\title{Remarks on cosmological bulk viscosity in different epochs}         

\author{Iver Brevik$^1$ and Ben David Normann}      
\affiliation{$^1$Department of Energy and Process Engineering, Norwegian University of Science and Technology, N-7491 Trondheim, Norway}
\date{\today}          

\begin{abstract}
The intention of this paper is mainly two-fold. \textit{First}, we point out a striking numerical  agreement between the bulk viscosity in the lepton era calculated by Husdal (2016) and our own calculations of the present-day bulk viscosity when the functional form is $\zeta\,\sim\sqrt{\rho}$. From a phenomenological point of view, we thus seem to have an ansatz for the viscosity which bridges the infancy of the Universe ($\sim 1$ s) with  the present. This can also be looked upon as a kind of symmetry between the early-time cosmology and the present-day cosmology: it is quite remarkable that  the kinetic theory-based bulk viscosity in the early universe and the experimentally-based bulk viscosity in the present universe can be covered by the same simple analytical formula.
\textit{Second}, we consider  the Kasner universe as a typical anisotropic model of Bianchi-type I, investigating whether this geometrical model is compatible with constant viscosity coefficients in the fluid. Perhaps surprisingly, the existence of a shear viscosity turns out to be incompatible with the Kasner model. By contrast, a bulk viscosity is non-problematic in the {\it isotropic} version of the model. In the special case of a Zel'dovich (stiff) fluid, the three equal exponents in the Kasner metric are even determined by the bulk viscosity  alone, independent of the value of the fluid energy density. We also give a brief comparison with some other recent approaches to viscous cosmology.

\end{abstract}
\maketitle

\bigskip
\section{Introduction}
\label{secintro}
\noindent Recent years have witnessed an increased interest in bulk-viscous properties in the cosmic fluid. From a hydrodynamicist's point of view it is almost surprising that this surge of interest has not occurred earlier. As is known, viscous effects are quite ubiquitous in ordinary hydrodynamics, and one should not expect the cosmic fluid to be an exception in that respect. There are in general two viscosity coefficients, the shear viscosity $\eta$ and the bulk viscosity $\zeta$, corresponding to first-order deviation from thermal equilibrium, although normally the shear viscosity is omitted in cosmology because of the assumption about spatial isotropy of the fluid. But what happens if one takes away the assumption about isotropy? We demonstrate that in the Kasner universe, shear viscosity seems unphysical. Hence we are left with bulk-viscous modifications to the equation of state. Moreover, we present in this paper a striking similarity between the results obtained for the bulk viscosity of the late Universe~\cite{normann16} based on a much-used phenomenological approach ($\zeta\,\sim\,\sqrt{\rho}$), and that calculated by Husdal and others (see below) for the early lepton-era.

We  will assume a spatially flat Friedmann-Lemaître-Robertson-Walker universe, where the  metric is
\begin{equation}
ds^2=-dt^2+a^2(t)d{\bf x}^2, \label{1}
\end{equation}
and the energy-momentum tensor   of the whole fluid is
\begin{equation}
T_{\mu\nu}=\rho U_\mu U_\nu+(p-\theta \zeta)h_{\mu\nu}, \label{2}
\end{equation}
with $h_{\mu\nu}=g_{\mu\nu}+U_\mu U_\nu$  as the projection tensor. The scalar expansion is $\theta={U^\mu}_{;\mu}=3H$. In comoving coordinates, the components of the fluid four-velocity are $U^0=1, U^i=0$.

The total energy density $\rho$ of the cosmic fluid is taken to be composed of several parts,
\begin{equation}
\rho=\rho_{\rm DE}+ \rho_{\rm DM}+\rho_{\rm B} +\rho_{\rm R}, \label{3}
\end{equation}
where subscripts ${\rm DE,\,DM,\,B}$ and ${\rm R}$ refer to to dark energy, dark matter, baryons and radiation, respectively. Experiments show that  the dark sector amounts to about 95\% of the total energy content \cite{planck18VI}. We will define  $\rho_{\rm M}$ as the sum of  the dark matter and  baryons,
\begin{equation}
\rho_{\rm M}= \rho_{\rm DM}+ \rho_{\rm B}. \label{4}
\end{equation}
The contribution from radiation is negligible. Defining the usual density parameters $\Omega_i=\rho_i/\rho_c$ with $\rho_c$ the critical density, we have
\begin{equation}
\Omega_{\rm DE}= \frac{\rho_{\rm DE}}{\rho_{\rm c}}, \quad \Omega_{\rm M}=\frac{\rho_{\rm M}}{\rho_{\rm c}}, \quad \rho_{\rm c}=\frac{3H^2}{8\pi\rm G}. \label{5}
\end{equation}
At present (subscript $0$) the Planck experiment finds  \cite[Table 2]{planck18VI} $\Omega_{\rm DE}= 0.6847,\, \Omega_{0\rm M} =0.3153$, summing up to unity. It is also useful to note that $H_0=67.36$ km~s$^{-1}$ Mpc$^{-1} = 2.20\times 10^{-18}~$s$^{-1}$, $\rho_{0c}=3H_0^2/8\pi G= 8.5\times 10^{-27}~$kg~m$^{-3}$.

Consider now  the Friedmann  equations
\begin{equation}
\theta^2=24\pi G \rho, \label{6}
\end{equation}
\begin{equation}
\dot{\theta} + \frac{1}{2}\theta^2= -12\pi G(p-\zeta \theta). \label{7}
\end{equation}
The obvious task is to figure out how to model the bulk viscosity\footnote{Although not a focus in the present work, we also remark that a positive, non-vanishing bulk viscosity will generate entropy~\cite{weinberg71}.} . One obvious choice---and the one made in this paper---is to model the bulk viscosity as a function of the total energy density,  $\zeta = \zeta(\rho)$. In Refs.~\cite{normann16,normann17} we advocated the power-law form
\begin{equation}
\zeta(\rho)=\zeta_0\left( \frac{\rho}{\rho_0}\right)^\lambda, \label{8}
\end{equation}
with $\lambda$ a constant. Preference, although not  a very strong one,  was given to the case $\lambda=1/2$, in agreement with several other investigators having compared with experiments.

\bigskip

Our motivations for undertaking the present investigation are the following:

\noindent 1. The phenomenological approach above was based  upon redshifts up to about $z=2.3$. It is of interest to make a big jump in the cosmological scale, back to the lepton-photon universe, characterized by temperatures between $T=10^{12}~$K and $10^{10}~$K, where the Universe was populated by photons, neutrinos, electrons, and their antiparticles. Also under such circumstances a bulk viscosity appears, explained in kinetic terms as a result of the imbalance between the free paths of neutrinos and the other particles. The maximum bulk viscosity occurs at the
time of neutrino decoupling,  $T=10^{10}~$K. We will base our analysis on the recent papers of Husdal et al. \cite{husdal16a,husdal16,husdal17}. Moreover, we will show that  a bold extrapolation of the formula (\ref{8}) with $\lambda=1/2$ back to this very early instant brings surprisingly good agreement with the kinetic-theory based result for $\zeta$. It becomes suggestive to assume that formula (\ref{8}) holds for very longer times back than what is so far justified from observations. In other words, there is apparently a kind of symmetry between the early-time cosmology and the present-day cosmology as far as the bulk viscosity is concerned. 

\noindent 2. Some effort ought to be made to clarify the reasons why there are apparently conflicting statements in the recent literature. The extensive analysis of Yang et al. \cite{yang19} favors the form $\zeta \propto \rho^m$ with $m \approx -0.4$, thus of a generalized Chaplygin form, and quite different from what we stated above. As one might expect, this discrepancy is rooted in differences in the initial formalism. We consider this theme in some detail in sect.~\ref{sec:yang}.

\noindent 3. The common omission of the shear viscosity in cosmology is not quite trivial, all the time that the shear viscosity is the dominant viscosity in ordinary fluid mechanics. One might suspect that even a slight anisotropy in the cosmic fluid could easily compensate for the bulk viscosity. We consider this point  in Sect.~\ref{sec:kasner}, choosing the anisotropic Kasner universe as an example. It actually turns out that the Kasner model is not easily compatible with a shear viscosity. On the other hand, the model admits a bulk viscosity without any problems, in the degenerate case of spatial isotropy.

\bigskip
Readers interested in review articles on viscous cosmology, may consult Refs.~\cite{weinberg71,gron90,brevikgron13,zimdahl96,brevikgron17}. By now the literature is rich with contributions on the topic of viscous cosmology. The contributions include investigations of the early universe~\cite{brevik15c,brevikTimoshkin16,bamba15,campo07}, the late universe~\cite{brevik11,brevik12,brevik15a}, the phantom divide~\cite{Frontiers,brevik06,velten13,brevik15b,nojiri05a,disconzi15}, models for the dark sector~\cite{elizalde14,Gorbunova,bamba12,brevik15d}, and others
~\cite{nojiri05,nojiri03,wang14,floerchinger15,brevik11a,barrow86,li10,cardenas15,paolis10,horn16}.

\section{Possible relationship to the bulk viscosity in the lepton-photon epoch}
\label{sec:possibleRel}
On the basis of $H(z)$ measured for different moderate values of $z$ ($0<z<2.3$) we estimated in Refs.~\cite{normann16,normann17} the present bulk viscosity $\zeta_0$ to lie in the interval
\begin{equation}
10^4 ~\rm{Pa~s} \leq \zeta_0 \leq 10^6 ~\rm{Pa~s}. \label{9}
\end{equation}
Although we refer the interested reader to the sources for detailed explanations, we will  in the following clarify under what assumptions these results were obtained. 
First of all, we made use of the ansatz (\ref{8}), testing three different values for the exponent $\lambda$, $\lambda= (0, 1/2, 1)$. The equation of state was assumed in the simple form $p=w\rho$ with $\rho$ a constant.
As mentioned earlier, the total energy density is assumed to be composed by a dark energy component $\rho_{\rm DE}$ and a matter component $\rho_{\rm M}=\rho_{\rm DM}+\rho_{\rm B}$ satisfying $\Omega_{\rm DE}+\Omega_{\rm M} =1$. Denoting the homogeneous solution corresponding to $\zeta =0$ by $\rho_{\rm h}(a)$, we can decompose
\begin{equation}
\rho_{\rm h}(a)=\sum_i \rho_{0i}a^{-3(w_i+1)}, \label{13}
\end{equation}
with $i=({\rm DE, M}), \, w_{\rm DE}=-1, \, w_M=0$. Moreover, for simplicity we assume that the viscosity can be associated with the fluid as a whole. Going beyond a phenomenological approach, this seems to imply that the viscosity directly or indirectly is sourced by the interaction of the various components of the phenomenological one-component fluid. Consider~\cite{colistete07,piattella11} for interesting discussions. Furthermore; a bulk viscosity associated with the overall fluid would be most accurate if the dark sector interacts with the constituents of the standard model. Associating the viscosity with the fluid as a whole enables us to write
\begin{equation}
\rho(a)=\rho_{\rm h}(a)[1+u(a)], \label{14}
\end{equation}
where $u(a)$ is determined from Friedmann's equations. It is useful to define the auxiliary quantity
\begin{equation}
B_0=12\pi {\rm G} \zeta_0, \label{15}
\end{equation}
whose value is about unity in astronomical units (note  the subscript $_0$ on $B$;  this subscript was not used in our previous works, but should have been there in order to indicate that this is a present-day value). We introduce the common notation $E=H/H_0$ and give the final formulas for the most actual option $\lambda=1/2$ only. Then,
\begin{equation}
u(z, B)=(1+z)^{-\frac{2B}{H_0}}, \label{16}
\end{equation}
\begin{equation}
E(z, B)=\sqrt{\Omega(z)}(1+z)^{-\frac{B}{H_0}}. \label{17}
\end{equation}
We have thus delineated the assumptions that led us to the result (\ref{9}), by comparison with the observations of $H(z)$. This is also largely in agreement with earlier investigators \cite{wang14,velten12}.

\bigskip

Let us now go back to the early Universe, the lepton-photon era, in which case the viscosity coefficients have to be calculated by kinetic theory. There are two factors that are important for the calculation, namely the state of the system (we assume it to be pure lepton-photon mixture), and then the transport equations for the fluid. The free mean paths for the neutrinos are much larger than those of the electromagnetically interacting particles, thus building up a temperature difference between the fluid components. The electromagnetic particles will cool somewhat faster than the neutrinos. Both coefficients $\eta$ and $\zeta$ can be evaluated via the Chapman-Enskog approximation \cite{hogeveen86,groot80,husdal16a,husdal16,husdal17}, here given for $\zeta$ only,
\begin{equation}
\zeta(T)=nk_{\rm B}T\sum_k a_k\alpha_k. \label{18}
\end{equation}
The $n$ is the particle density, $a_k$ are the coefficients for particle $k$ for the linearized relativistic Boltzmann equation, and $\alpha_k$ are known state parameters. As mentioned above, the most significant instant is that of neutrino decoupling, $T=10^{10}~$K, at which  Husdal obtains \cite{husdal16}
\begin{equation}
\zeta=1.26\times 10^{22}~{\rm Pa~s}, \quad \eta=1.0\times 10^{25}~\rm{Pa~s}. \label{19}
\end{equation}
This value of $\zeta$ can be compared with that obtained from the expression (\ref{8}) extrapolated back to the instant of neutrino decoupling. The relation between $\rho$ and $T$ in the early Universe is \cite{husdal17} (in geometric units)
\begin{equation}
\rho=\frac{\pi^2}{30}g_*(T)(k_BT)^4, \label{20}
\end{equation}
with $g_*$ denoting the effective degrees of freedom at temperature $T$. Table 1 shows the resulting estimates for $\zeta$ for the three actual parameter values $\lambda =(0, 1/2, 1)$. What is apparent is that,  by choosing $\lambda =1/2$, and also by taking $\zeta$ equal to  the logarithmic mean of the interval (\ref{9}), i.e.
\begin{equation}
\zeta_0=10^5~\rm{ Pa~s}, \label{21}
\end{equation}
we find practically the same value of $\zeta$ as with \eqref{8}.

\begin{table}[h]
\centering
\begin{tabular}{lcccccr}
\toprule
\multicolumn{7}{c}{\textbf{Table showing $\zeta$ estimates}} \\
\hline
$\lambda$ $[-]$&&$\zeta_0$  $[{\rm Pa\,s]}$&&$\Omega_0$&& $\zeta$ from Eq. \eqref{8} $[{\rm Pa\,s]}$\\
\hline
0&&$\forall\,\zeta_0$&&(0,1]&&$\zeta\,=\,\zeta_0$\\
\hline
$1/2$&&$\zeta_0=10^{4}$&&[0.1,1]&&$\sim10^{21}$\\
$1/2$&&$\zeta_0=10^{5}$&&[0.1,1]&&$\sim 10^{22}$\\
$1/2$&&$\zeta_0=10^{6}$&&[0.1,1]&&$\sim 10^{23}$\\
\hline
$1$&&$\zeta_0=10^{4}$&&[0.9,1]&&$\sim 10^{38}$\\
$1$&&$\zeta_0=10^{5}$&&[0.9,1]&&$\sim 10^{39}$\\
$1$&&$\zeta_0=10^{6}$&&[0.9,1]&&$\sim 10^{40}$\\
\hline
\end{tabular}
\caption{The table should be read horizontally and shows the numerical values of $\zeta$ calculated by Eq.~(\ref{8}) 
 for different parameter value inputs $\{\lambda,\zeta_0,\Omega_0\}$. The parameter values given in the three first columns are used as input when calculating the values for $\zeta$ in the rightmost column.}
\label{tab:Visc}
\end{table}

We find this coincidence striking. Of course, it is not a proof for the extended applicability of the formula (\ref{8}), but we think it deserves attention. It suggests that the formula can be used beyond the interval where it was originally constructed.  Also, it is notable that the parameter value $\lambda=1/2$ turns out to be the favorable choice, in agreement with other analyses as mentioned earlier.
 
\section{Discussion}
\noindent In ending this section, we shall show that the results obtained in this paper are actually not surprising at all, when considered from a phenomenological point of view, where the fluid has an overall viscosity. Consider the Bianchi identity for the overall cosmological fluid $\rho$;
\begin{equation}
    \dot{\rho}+\theta(1+w)\rho-\zeta\theta^2=0.
\end{equation}
This equation may be rewritten as
\begin{equation}
\dot{\rho}=-\theta(1+w(1-r))\rho,
\end{equation}
where we have defined
\begin{equation}
\label{r}
   r\,\equiv\, \frac{\theta}{w\rho}\zeta
\end{equation}
Thus $r$ is the ratio between the viscous pressure $\zeta\theta$ and the equilibrium pressure $p=w\rho$. In the Eckart-formalism the viscous pressure exerted by $\zeta$ should remain a first-order modification to the equilibrium pressure $w\rho$ throughout the history of the Universe. This translates into the requirement that $0\leq r\ll1$. Then, under the weak assumption that $\zeta$ is a monotonic function, we find
\begin{equation}
\label{cond}
   r\approx\rm const.
\end{equation}
Detailed analysis of the dynamical behaviour of the cosmological fluid could of course reveal small variations in the ratio $r$, so this relationship represents an approximation. Moreover, the variation in $r$ could very well prove to be important in a variety of contexts\footnote{such as that of understanding the (microscopic) mechanism that gives rise to the viscosity, or structure formation.}, but in this phenomenological analysis we are nevertheless more interested in large-scale variations. Thus putting $r=\textrm{const.}$ translates into $\zeta\sim\rho/\theta$. For a one-component fluid $\rho$ this is, by application of the first Friedmann equation just the same as ansatz~\eqref{8}. As such, our result is trivial, yet obviously worth mentioning, considering the wide application of other functional forms for the viscosity, also in the case of one-component fluids. The fact that the results of Husdal actually agree with our own calculations, as inferred from supernova observations, shows that our theoretical prejudice seems to be confirmed by observation. This is, of course, not a trivial point. In more general terms, consider a viscosity $\zeta\sim\rho^m$ and an equilibrium pressure $w\rho$. Then the ratio $r$ defined in
~\eqref{r} becomes
\begin{equation}
\label{condGeneral}
    r\,\sim\,\rho^{m-1/2}.
\end{equation}
The choice $m=1/2$ now gives the condition~\eqref{cond}. One may also observe that any $m<1/2$ will cause the ratio to grow if $\rho$ decreases. Consequently, unless $\rho$ asymptotically approaches a constant value, the ratio must therefore eventually grow out of bounds of the first-order thermodynamic (Eckart) formalism.

\subsection*{Comparison with the result $\zeta_{\rm D}\sim\rho^{-0.4}$ obtained by Yang et al.~\cite{yang19} }
\label{sec:yang}
As mentioned in the introduction, Yang et al. suggest a model where the effective pressure $\rho_{\rm eff}$ of the dark fluid $\rho_{\rm D}=\rho_{\rm DE}+\rho_{\rm DM}$ is
\begin{equation}
    p_{\rm eff}=w\rho_{\rm D}+\sqrt{3}\alpha\rho\cdot\rho_{\rm D}^{m-1},
\end{equation}
where $w$ is the equation-of-state parameter, $\alpha$ and $m$ are parameters of the theory and $\rho$ is the total energy density. The above equation makes sense only if $p_{\rm eff}$ and $w$ is (respectively) the effective pressure and equation-of-state parameter for the unified dark fluid only, and not the overall fluid $\rho$. We therefor henceforth make this assumption. Thus interpreting their results, we reach the conclusion that they work with a different theory from our own. While they attribute the viscosity to the dark fluid only, we attribute the viscosity to the overall fluid. Perhaps more importantly, we fixed the dark-matter and dark-energy components at present when obtaining our estimates for the present-day viscosity. As far as we understand, this is different from Yang et al. who (more appropriately) avoided such an a-priori fixing of parameters. To sum up, the discrepancy derives from the difference in the theories. Without further examination, one may not conclude that the two approaches are in disagreement, per say. We shall repeat from the preceding subsection, however, that $r$ in Eq.~\eqref{condGeneral} will eventually grow out of bounds for $m=-0.4$, since $-0.4<1/2$. Hence, a viscosity $\zeta\sim\rho^{-0.4}$ cannot be considered as a bulk-viscous modification in the ordinary sense, unless $\lim_{t\rightarrow\infty}\rho={\rm const.\,\neq\,0}$. As a general dynamical modification to a homogeneous equation of state, the results of Yang et al. seem however to be valid.

\section{The viscous Kasner universe}
\label{sec:kasner}
\noindent It is of interest to consider the anisotropic universe. The reason why the Universe is usually considered to be spatially isotropic, is that observation strongly indicates such behavior. However, there is also  evidence, in cosmology as well as in ordinary fluid mechanics, that the shear viscosity grossly dominates over the bulk viscosity in magnitude. Thus it might be possible that the combination of a slight anisotropy with a dominant shear viscosity leads to physically detectable consequences after all. This is the motivation for the analysis in the present section. We will focus on the anisotropic Kasner universe, as a typical example of an anisotropic space. It  belongs to the Bianchi-type I. An introduction to this model can be found, for instance, in Ref.~\cite{landau75}.

Consider the Kasner universe:
\begin{equation}
 ds^2=-dt^2+t^{2p_1}dx^2+t^{2p_2}dy^2+t^{2p_3}dz^2, \label{metric}
 \end{equation}
 where the three $p_i$ are constants, in the original formulation which refers to a vacuum.
Einstein's equations can be written
\begin{equation}
 R_{\mu\nu}=8\pi G(T_{\mu\nu}-\frac{1}{2}g_{\mu\nu}T^\alpha_\alpha), \label{40}
 \end{equation}
and the non-vanishing Christoffel symbols are (no sum over $i$):
\begin{equation}
 \Gamma^0_{ii}=p_it^{2p_i-1}, \quad \Gamma_{i0}^i =\Gamma_{0i}^i=\frac{p_i}{t}. 
\end{equation}
Allowing for both shear and bulk viscosities, we write the energy-momentum tensor as
\begin{equation}
    T_{\mu\nu}=\rho U_{\mu}U_{\nu} +(p-\zeta \theta)h_{\mu\nu}-2\eta \sigma_{\mu\nu},
\end{equation}
where the scalar expansion $\theta=\theta^\alpha_\alpha ={U^\alpha}_{;\alpha}$ is the trace of the expansion tensor
\begin{equation}
    \theta_{\mu\nu}=\frac{1}{2}(U_{\mu ; \alpha}h_\nu^\alpha +U_{\nu; \alpha}h_\mu^\alpha ),
\end{equation}
and $\sigma_{\mu\nu}$ is the shear tensor
\begin{equation}
\sigma_{\mu\nu}=\theta_{\mu\nu}-\frac{1}{3}h_{\mu\nu}\theta.
\end{equation}
Defining the numbers $S$ and $Q$ as
\begin{equation}
    S=\sum_{i=1}^3p_i, \quad Q=\sum_{i=1}^3p_i^2,
\end{equation}
we can then write
\begin{equation}
    \theta= \frac{S}{t}, \quad \sigma^2= \frac{1}{2}\sigma_{\mu\nu}\sigma^{\mu\nu}=-\frac{1}{2t^2}(\frac{1}{3}S^2-Q). \label{46}
\end{equation}
With $R_i=t^{p_i}$ being the expansion factors of the metric, the directional Hubble parameters become $H_i=\dot{R}_i/R_i=p_i/t$, and the average Hubble parameter becomes
\begin{equation}
    H=\frac{1}{3}\sum_{i=1}^3H_i=\frac{S}{3t} = \frac{\theta}{3}.
\end{equation}

Let now $w=$constant be the thermodynamic parameter,
\begin{equation}
    p=w\rho.
\end{equation}
With  $\kappa=8\pi G$ we can then write the Einstein equations (\ref{40}) as
 \[    S-Q+\frac{3}{2}\kappa t \zeta S =\frac{1}{2}\kappa t^2(1+3w)\rho, \]
 \begin{equation}
     p_i(1-S-2\kappa t\eta)+\frac{1}{2}\kappa t\left( \zeta+\frac{4}{3}\eta \right)S=\frac{1}{2}\kappa t^2(w-1)\rho. \label{47}
 \end{equation}
 The basic formalism given here is as in our earlier works \cite{brevik97,brevik00}. We are thus considering an {\it isotropic} fluid in an {\it anisotropic} space\footnote{What we mean by this is that the same fluid in an isotropic background, would not possess shear viscosity.}. The fluid itself is modeled as a usual fluid with density $\rho$ and scalar pressure $p$. Our model is thus different from one in which  the fluid  is taken to have anisotropic properties; cf., for instance, Ref.~\cite{cadoni20}. We will assume that the physical properties of the fluid are given at some initial time called $t_{\rm in}$, and investigate if these initial conditions are compatible with constant values of  $S$ and $Q$. In the vacuum case (no fluid at all), one has $S=Q=1$ \cite{landau75}. We expect the Kasner model to be appropriate for the early Universe, and will naturally choose the instant of neutrino decoupling, $T_{\rm in}= 10^{10}~$K,  $t_{\rm in}=1~$s, as mentioned above. 
 
 We consider the development of the fluid from $t=t_{\rm in}$ onwards; the initial energy and pressure being $\rho_{\rm in}$ and $p_{\rm in}$. It is notable that the governing equations (\ref{47}) actually fix the later time dependence to be
 \begin{equation}
     \zeta=\zeta_{\rm in}\left( \frac{t}{t_{\rm in}}\right)^{-1}, \quad \eta= \eta_{\rm in}\left( \frac{t}{t_{\rm in}}\right)^{-1}, \label{49}
 \end{equation}
\begin{equation}
    \rho=\rho_{\rm in}\left( \frac{t}{t_{\rm in}}\right)^{-2}, \quad p=p_{\rm in}\left( \frac{t}{t_{\rm in}}\right)^{-2}, 
\end{equation}
whereby we obtain a time-independent set of equations,
\[ S-Q+\frac{3}{2}\kappa t_{\rm in}\zeta_{\rm in}S= \frac{1}{2}\kappa t_{\rm in}^2(1+3w)\rho_{\rm in}, \]
\begin{equation}
p_i(1-S-2\kappa t_{\rm in}\eta_{\rm in})+\frac{1}{2}\kappa t_{\rm in}( \zeta_{\rm in}+\frac{4}{3}\eta_{\rm in})S= -\frac{1}{2}\kappa t_{\rm in}^2(1-w)\rho_{\rm in}, \label{50}
\end{equation}
(note that in fundamental units where the basic unit is $[t]=~$cm, one has $[\rho]=[p]=~$cm$^{-4}$, $[\zeta]=~$cm$^{-3}$). It is also convenient to note the following equations derived from those above,
\begin{equation}
    2S(S-1)=3\kappa t_{\rm in} \zeta_{\rm in}S+3\kappa t_{\rm in}^2(1-w)\rho_{\rm in}, \label{51}
\end{equation}
\begin{equation}
   S^2-Q=2\kappa t_{\rm in}^2\rho_{\rm in}. \label{53}
\end{equation}
The basic formalism outlined so far is essentially as in our earlier papers \cite{brevik97,brevik00}. We assume now that the physical quantities $\{\rho_{\rm in},p_{\rm in},\zeta_{\rm in},\eta_{\rm in}\}$ are given at $t=t_{\rm in}$, and investigate if these initial conditions lead to acceptable values for the coefficients $p_i$ in an anisotropic Kasner universe. From Eq.~(\ref{50}) it follows that if the three $p_i$ are to be unequal, the multiplying factor of $p_i$ has to be zero,
\begin{equation}
    S=1-2\kappa t_{\rm in}\eta_{\rm in}. \label{54}
\end{equation}
This is thus an algebraic restriction on the $p_i$. If the fluid is nonviscous, then $S=1$, in accordance with the original Kasner model in vacuum. Once $S$ is known, $Q$ follows at once from Eq.~(\ref{53}) as
\begin{equation}
    Q=S^2-2\kappa t_{\rm in}^2\rho_{\rm in}. \label{55}
\end{equation}
Note that the bulk viscosity does not appear in the two last equations. This is as should be expected physically: anisotropy is caused by the shear only.

 Now going over to dimensional units we first note the useful relations $\kappa = 8\pi G/c^2 = 1.866 \times 10^{-26}~$m~kg$^{-1}$, 1 MeV$^4=2.085\times 10^{25}~$J~m$^{-3}$. In the radiation dominated era
 \begin{equation}
  a(t)=2.2\times 10^{-10}~t^{1/2}, \quad T(t)= 10^{10}~t^{-1/2}~\rm{K}.
 \end{equation}
 With $\eta_{\rm in}=1.0\times 10^{25}~$Pa~s we then obtain from Eq.~(\ref{54})
\begin{equation}
    S=0.627.
\end{equation}
The energy density in this region can be roughly estimated from $\rho_{\rm in}c^2=a_{\rm rad}T^4$, where $a_{\rm rad}$ is the radiation constant
\begin{equation}
a_{\rm  rad}=\frac{\pi^2 k_B^4} {15 \hbar^3 c^3}
  =7.56 \times 10^{-16}~{\rm J}~{\rm}m^{-3}~K^{-4}.  
\end{equation}
Here the degeneracy factor is omitted (cf., for instance, Ref.~\cite{brevik15c}). Then,  $\rho_{\rm in} \sim 10^{25}~$J~m$^{-3}$. Alternatively, we may use the equation $H_{\rm in}^2=(\kappa c^2/3)\rho_{\rm in}$ to get somewhat more accurately
\begin{equation}
\rho_{\rm in}=4.47\times 10^8~{\rm kg~m}^{-3},
\end{equation}
or $\rho_{\rm in} =4.02\times  10^{25}~$J~m$^{-3}$. With the latter values we calculate
\begin{equation}
    Q=1-2\kappa c^2t_{\rm in}^2\rho_{\rm in}=-1.11.
\end{equation}
This is a non-acceptable result, as $Q$ is a sum of quadratic numbers. We conclude that the 
Kasner model  does not appear to be compatible with a shear viscosity.

There is also another reason why anisotropy ($p_i=0$) is problematic in the Kasner model. From Eq.~(\ref{47}) it follows that if $w<1$, which is the case for the usual fluids, the combination $(\zeta +4\eta/3)$ becomes negative. Although negative viscosities are occasionally considered in cosmology (cf. for instance, Ref.~\cite{brevik13}), such a case is physically not very natural. This point was first noticed by Cataldo and Campo \cite{cataldo00}. 

It is  worthwhile to notice that the {\it isotropic} Kasner geometry easily allows for a viscosity. That means, only the bulk viscosity comes into question, as this viscosity concept goes along with spatial isotropy. Let us consider this case in some more detail, defining $a$ by  $p_1=p_2=p_3 \equiv a$. From Eq.~(\ref{50}) we obtain, still in dimensional units,
\begin{equation}
    a= \frac{1}{6}\left[ 1+\frac{3}{2}\kappa t_{\rm in}\zeta_{\rm in} +\sqrt{\left( 1+\frac{3}{2}\kappa t_{\rm in}\zeta_{\rm in}\right)^2+6\kappa c^2t_{\rm in}^2(1-w)\rho_{\rm in}} \right].
\end{equation}
This equation shows that if the physical quantities $\rho_{\rm in}$ and $p_{\rm in}$, as well as the bulk viscosity $\zeta_{\rm in}$, are given at $t=t_{\rm in}$ then the isotropic version of the Kasner metric  (\ref{metric}) is determined.

There is finally one exceptional case that should be noticed, namely a Zel'dovich fluid for which $w=1$, the velocity of sound being equal to $c$. In that case,
\begin{equation}
a= \frac{1}{3}\left[ 1+\frac{3}{2}\kappa t_{\rm in}\zeta_{\rm in} \right].  \label{isotropy}
\end{equation}
The metric  turns  out to be determined by the bulk viscosity as the only physical parameter; the actual value of $\rho_{\rm in}$
being irrelevant. If the fluid is nonviscous, $\zeta_{\rm in}=0$, then $a=1/3$. It corresponds to the metric coefficients in Eq.~(\ref{metric}) being $t^{2a} = t^{2/3}$.

\section{Conclusion}
In this work we have pointed out an agreement between the bulk-viscosity calculated by Husdal et al. \cite{husdal16a,husdal16,husdal17} to be present in the lepton-photon era and the present-day phenomenological viscosity calculated by ourselves in previous works based on the evolution $\zeta\sim\sqrt{\rho}\sim H$ of the viscosity in isotropic cosmology.  As mentioned above, this points towards a symmetry between early-time and present-time viscous cosmology. It is  instructive to note that viscosities such as $\zeta\sim\rho$ or constant $\zeta$ do \textit{not} provide the same agreement.

Our analysis of the Kasner model showed the inapplicability of the shear viscosity concept in the anisotropic version of this model (meaning that at  least two of the three exponents $p_i$ in the metric (\ref{metric}) are unequal). The isotropic Kasner model is however easily compatible with a bulk viscosity. Then the Kasner model itself becomes isotropic, so that $p_1=p_2= p_3$. It is noteworthy that for the special case of a Zel'dovich fluid the Kasner exponents are determined exclusively by the bulk viscosity alone; cf. Eq. (\ref{isotropy}).


\newpage



\end{document}